# High spatial resolution dosimetry with uncertainty analysis using Raman micro-spectroscopy readout of radiochromic films

Connor Mcnairn[1], Iymad Mansour[1], Bryan Muir[2], Rowan M. Thomson[1] and Sangeeta Murugkar[1]

[1]Department of Physics, Carleton University, 1125 Colonel By Drive, Ottawa, Ontario K1S 5B6, Canada

[2]Metrology Research Centre, National Research Council of Canada, Ottawa, ON, K1A 0R6 Canada

*Corresponding author: smurugkar@physics.carleton.ca

**Purpose:** The purpose of this work is to develop a new approach for high spatial resolution dosimetry based on Raman micro-spectroscopy scanning of radiochromic film (RCF). The goal is to generate dose calibration curves over an extended dose range from 0-50 Gy and with improved sensitivity to low (< 2 Gy) doses, in addition to evaluating the uncertainties in dose estimation associated with the calibration curves.

**Methods**: Samples of RCF (EBT3) were irradiated at a broad dose-range of 0.03 Gy-50 Gy using an Elekta Synergy clinical linear accelerator. Raman spectra were acquired with a custom-built Raman micro-spectroscopy setup involving a 500 mW, multimode 785 nm laser focused to a lateral spot diameter of 30 µm on the RCF. The depth of focus of 34 µm enabled the concurrent collection of Raman spectra from the RCF active layer and the polyester laminate. The pre-processed Raman spectra were normalized to the intensity of the 1614 cm$^{-1}$ Raman peak from the polyester laminate that was unaltered by radiation. The mean intensities and the corresponding standard deviation of the active layer Raman peaks at 696, 1445 and 2060 cm$^{-1}$ were determined for the 150 × 100 µm$^2$ scan area per dose value. This was used to generate three calibration curves that enabled the conversion of the measured Raman intensity to dose values. The experimental, fitting and total dose uncertainty was determined across the entire dose range for the dosimetry system of Raman micro-spectroscopy and RCF.

**Results:** In contrast to previous work that investigated the Raman response of RCFs using different methods, high resolution in the dose response of the RCF, even down to 0.03 Gy, was obtained in this study.

The dynamic range of the calibration curves based on all three Raman peaks in the RCF extended up to 50 Gy with no saturation. At a spatial resolution of $30 \times 30$ μm$^2$, the total uncertainty in estimating dose in the 0.5 Gy to 50 Gy dose range was [6 − 9]% for all three Raman calibration curves. This consisted of the experimental uncertainty of [5 − 8]%, and the fitting uncertainty of [2.5 - 4.5]%. The main contribution to the experimental uncertainty was determined to be from the scan area inhomogeneity which can be readily reduced in future experiments. The fitting uncertainty could be reduced by performing Raman measurements on RCF samples at further intermediate dose values in the high and low dose range.

**Conclusions:** The high spatial resolution experimental dosimetry technique based on Raman micro-spectroscopy and RCF presented here, could become potentially useful for applications in microdosimetry to produce meaningful dose-estimates in cellular targets, as well as for applications based on small field dosimetry that involve high dose gradients.

## 1. INTRODUCTION

Raman spectroscopy is an optical technique that uses a laser beam to perform non-destructive, real-time measurements of the composition of chemical mixtures without the use of an external contrast agent (label-free). It involves the inelastic scattering of laser light due to vibrations of molecular bonds and provides a "chemical fingerprint" of the specimen[1] allowing quantification of the relative proportion of molecular constituents such as lipids, proteins, DNA and specific amino acids. Combined with machine learning techniques, Raman spectroscopy has been applied to cells, tissue and biofluids demonstrating its potential as a powerful biomedical diagnostic technique.[2] It has been applied to detect and diagnose cancer *in vitro*, *ex vivo* and *in vivo*[3] as well as to investigate the effects of treatment.[4] In particular, Raman spectroscopy has been successfully applied to detect distinct biochemical response of cells and tissue exposed to ionizing radiation. Raman micro-spectroscopic studies with a spatial resolution of 10 μm investigated the radiobiological response in human prostate, breast and lung tumor cells *in vitro,* irradiated at relatively high doses up to 50 Gy.[5] Radiation-induced accumulation of intracellular glycogen was detected *ex vivo* using Raman micro-spectroscopy of tumor xenografts irradiated *in vivo* at a high radiation dose of 5 Gy and 15 Gy.[6] High-resolution confocal Raman spectroscopy with a spatial resolution of 1 μm has been applied to investigate the sub-cellular response of human lens epithelial cells,[7] and human keratinocytes[8] to ultra-low doses (from a few mGy up to 1 Gy) of ionizing radiation.

These studies indicate the potential for future application of Raman spectroscopy-based biomarkers in monitoring and predicting radiation response in individuals, and in understanding health risks from low dose radiation exposures. However, there are a couple of major challenges in such investigations. These studies typically involve mapping the Raman response from several microscopic locations inside a ~100 μm$^2$ area, and assume radiation dose (energy) to be deposited in a macroscopic, homogeneous manner. However computational dosimetry approaches involving Monte Carlo simulations have shown that there is considerable variation in energy deposition for low doses and microscopic cellular targets due to the stochastic nature of the energy deposition.[9] High spatial resolution experimental dosimetry[10] is thus required to accurately evaluate the energy deposition within microscopic volumes. The second challenge is that the spatial mapping of the Raman response in cells and tissue is confounded by the inherent intra- and inter-tumor or inter-cell heterogeneity or by non-targeted bystander effects.[6,8] Contributions from such confounding factors potentially can be removed by using a high spatial resolution experimental dosimetry system that involves Raman mapping of the radiation induced response of an inorganic substrate.

High spatial resolution dosimetry systems are actively being developed based on, for example, silicon, diamond, optical fiber, radiochromic film (RCF) and fluorescent nuclear track detectors.[11] RCFs have been used for two-dimensional radiation dosimetry over the last couple of decades.[12,13] Owing to their tissue-equivalence, energy-independent response and high spatial resolution, RCFs are widely used for radiotherapy treatment verification and quality assurance.[14,15] RCFs contain an active layer of diacetylene monomers that polymerizes to form polydiacetylene (PDA) in response to incident ionizing radiation.[16] The resulting change in the optical density is typically read out using transmitted or reflected visible light. The resolution of the RCF dosimetry system varies depending on the RCFs and the instrumentation used to read out the change in their optical density. The most common method of read out uses flatbed optical scanners and different flatbed optical scanning systems employed for reading out the same type of RCF can provide a spatial resolution of 350 µm[17] or 85 µm.[18] Recent work demonstrated a spatial resolution of $1 \times 1000$ µm$^2$ when using an optical microscope,[19] or 20 µm when using a microdensitometer[20] to replace the flatbed optical scanner for read out. Alternate optical readout techniques that provide high resolution have also been recently explored for RCFs. For example, a fiber-optic probe provided real-time measurements of the changes in optical transmission of the RCF with dose, across a broadband wavelength range.[21]

Confocal Raman spectroscopy has been applied to measure chemical changes in RCFs in response to the incident ionizing radiation.[16,22-24] This work underscores the potential of RCFs for micrometer-resolved dosimetry using Raman spectroscopy read-out. For Raman spectroscopy to become a robust technique for RCF dosimetry, it is critical to use standardized Raman data that are free of spectral measurement artifacts not related to the effects of the ionizing radiation. In addition to background correction, data pre-processing techniques such as correction of the instrument-dependent Raman intensity response and data normalization are crucial steps that must be included in the Raman data standardization procedure.[25,26] Moreover a Raman spectroscopy-based two-dimensional dosimetry system must not only provide the calibration curve but must also provide an estimate of the uncertainty in the value of dose predicted.[12] In this work, we demonstrate a novel measurement technique employing high-resolution Raman micro-spectroscopy instead of confocal Raman spectroscopy used in previous work. This enables concurrent collection of the Raman signal of the PDA in the RCF active layer and the overlaying polyester layer. After applying appropriate data standardization techniques that include normalizing the radiation induced response of the active layer with respect to the response of the polyester laminate which is unaltered by radiation, we demonstrate for the first time calibration curves that are almost linear in the low dose range providing improved sensitivity to low doses, in addition to an extended range for the dose-response curves compared to earlier work. Moreover, we provide a detailed characterization of the uncertainty in the estimation of the delivered dose at microscopic spatial resolution for the two-dimensional dosimetry system of the RCF and Raman micro-spectroscopy read-out.

## 2. MATERIALS AND METHODS

### 2.1. Radiochromic films

GAFChromic[TM] EBT3 radiochromic film (Ashland Specialty Ingredients, NJ, U.S.A.) was utilized in this study. The detailed specifications for these films are available online.[27] Briefly, the film consists of a 28 μm thick active layer of diacetylene monomers, sandwiched between two 125 μm thick polyester substrates. The dynamic dose range is 0.1 Gy to 20 Gy, with the optimum dose range specified as 0.2 Gy to 10 Gy. A single sheet of RCF (Lot #: 05091902) was cut into $5 \times 2$ cm$^2$ pieces, with one corner of each piece being marked to maintain the same film orientation for the radiation exposure and Raman spectral measurements. RCF samples were carefully handled with nitrile gloves and kept in darkness or low-light levels to minimize artifacts in the Raman spectra.

### 2.2. Radiation Exposure

Irradiations of RCF samples were performed using the Elekta Synergy clinical linear accelerator at National Research Council Canada (NRC) as shown in Fig. 1, with the nominal photon beam energy of 6 MV. The beam direction was perpendicular to the surface of the film. A single $5 \times 2$ cm$^2$ piece of RCF was placed at a depth of 5 cm at the center of a $30 \times 30$ cm$^2$ Virtual Water[TM] (Med-Tec) phantom, with approximately 15 cm of virtual water downstream to provide sufficient backscatter. The source-to-surface distance (SSD) to the front of the Virtual Water[TM] (Med-Tec) phantom was set to 100 cm using a calibrated mechanical pointer and the field size was set to $10 \times 10$ cm$^2$ at the phantom surface. A crosshair at the center of the phantom was aligned with the mechanical pointer and verified with the linac light field. The delivered dose per Monitor Unit (MU), measured using a calibrated NRC secondary standard reference chamber (type PTW-30013) connected to a Keithley 6517A electrometer, was measured both before and after RCF irradiations to ensure any drift in linac delivery is considered. Individual pieces of RCF were irradiated to 0.03, 0.15, 0.25, 0.5, 0.75, 1.0, 1.25, 1.50, 1.75, 2.0, 4.0, 6.0, 8.0, 10, 20, 30, and 50 Gy. The uncertainty on delivered dose in this irradiation geometry at NRC is 0.6% (k = 1) for doses greater than 0.5 Gy.[28-30] For doses less than 0.5 Gy, due to internal measurements of linac start-up effects, the uncertainty on the delivered dose increases from 0.6% to approximately 2.6%.

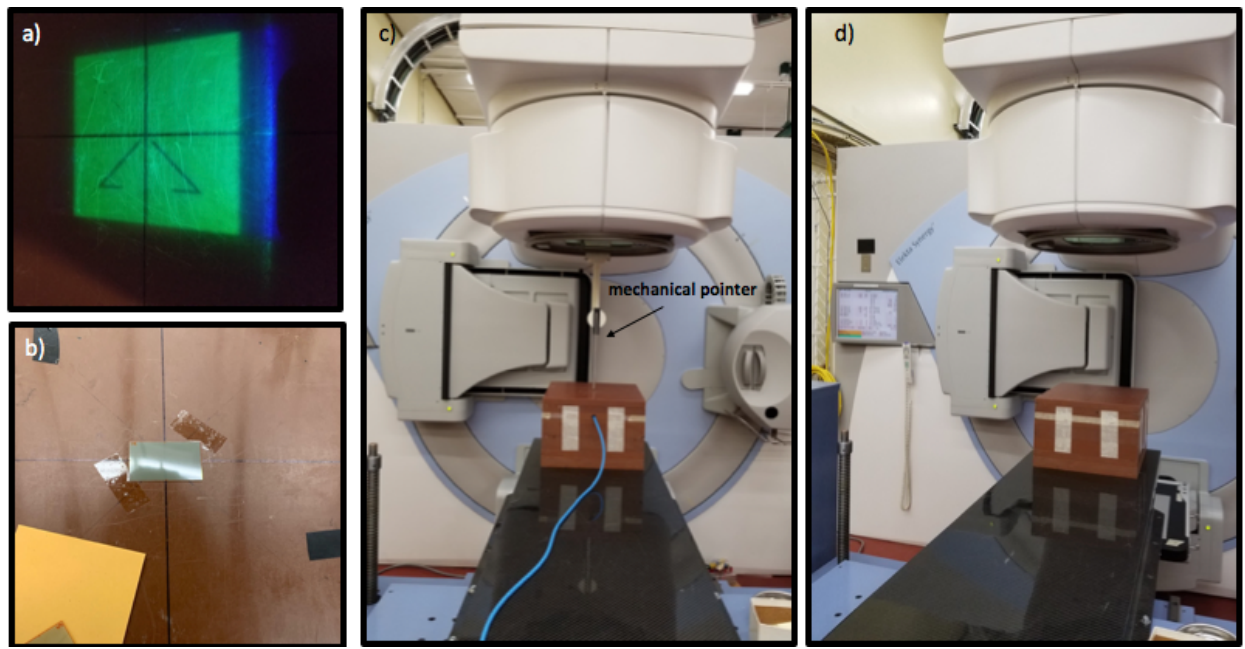

Fig. 1. (a) Linac light field used to place virtual water phantom in center of field. Crosshairs on the virtual water are used to ensure placement of the phantom in the center of the radiation field. (b) Placement of RCF in solid water phantom aligned with the center of the virtual water phantom. (c) Virtual water phantom with ionization chamber. (d) Virtual water phantom with RCF.

## 2.3. Raman micro-spectroscopy setup

Figure 2 (a) illustrates a schematic of the custom-built Raman micro-spectroscopy setup[31] used in this study. It consists of a multi-mode 785 nm diode laser (Ondax, CA, U.S.A.) with a variable output power of 0 - 500 mW coupled into a multi-mode fiber with a 100 μm core, and numerical aperture (NA) of 0.22. The laser light is collimated and passed through a long-pass dichroic mirror followed by a 40X, 0.8 NA water immersion microscope objective (Olympus, ON, Canada). This creates a multi-mode laser spot with a lateral diameter of ~30 μm and a depth of focus of 34 μm. As seen in Fig. 2 (b), Raman signal is thus collected from the active layer of the RCF along with parts of the overlaying polyester layer. The RCF sample is placed in the laser focal volume by adjusting its position by means of a three-axis automated stage (FTP 2000, ASI Inc., U.S.A) with a PZ-2000 piezo z-axis stage and controller having a resolution of ~220 nm. Raman scattered light from the sample is collected by the same microscope objective and passes through the dichroic mirror and two long-pass edge filters (Iridian, ON, Canada) to eliminate the 785 nm excitation laser light. A 300 μm core multimode fiber (Thorlabs, NJ, U.S.A.) delivers the Raman signal to a compact spectrometer (HyperFlux U1, Tornado Spectral Systems, ON, Canada) utilizing a 1200 line/mm grating and a proprietary high-throughput virtual slit.[32] Spectra were measured in the 170 to 2400 cm$^{-1}$ range, with a spectral resolution of 4 cm$^{-1}$.

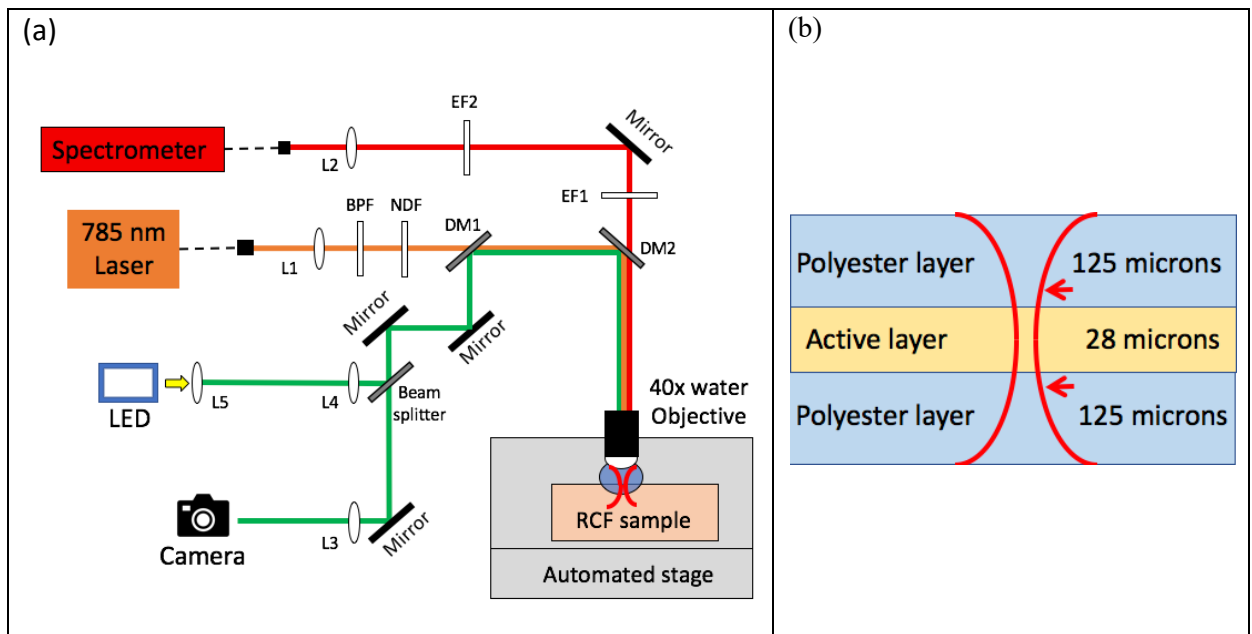

Fig. 2 (a) Schematic for Raman micro-spectroscopy setup to measure RCF sample. L=Lens; DM = Dichroic mirror; BPF = Bandpass filter; NDF = Neutral Density Filter; EF = Edge Filter (b) Schematic of the RCF cross-section shown with the incident laser focus spot. The arrows indicate the depth of focus of 34 μm of the laser spot.

## 2.4. Raman read-out operation

The Raman measurements reported here were performed two months after the irradiations in a climate-controlled (20.5 +/- 0.5 °C) laboratory environment. Polymerization of the active layer of the RCFs were confirmed to have stabilized through repeated Raman measurements. RCF samples were held in place on a custom-made solid aluminum slide to ensure flatness of the RCF samples. The position along the z-axis was adjusted by first obtaining the focused back-reflected laser spot from the RCF surface in the bright-field camera image and then vertically translating the stage to maximize the Raman signal from the active layer of the RCF. The z-axis stage with a sub-micron resolution provided the consistency in placing the 28 µm-thick active layer within the laser depth of focus of 34 µm. Raman spectra were obtained by moving the RCF sample horizontally through the laser focus in a raster scan pattern over an area of $150 \times 100$ µm$^2$ also referred to as the region of interest (ROI) in this work. This is similar to the 'Raman mapping' technique that was introduced earlier by Mirza and co-workers.[22,24] Preliminary measurements of the unirradiated (control) RCF samples revealed a nonlinear intensity response as a function of the focused laser power at the RCF sample. Hence a neutral density filter was used to lower the laser power at the sample in order to avoid any artifacts in the Raman intensity response due to photodamage of the RCFs. An optimum laser excitation power was determined to be 20 mW for which the active layer peak did not show increase in the Raman intensity with time. The data collection time per spectrum was set to 200 ms. When measuring sample sets with doses greater than ~20 Gy, the collection time was reduced to 90 ms to avoid oversaturating the spectrometer. The sample scanning step size of the raster scan was determined to be ~0.55 and ~0.25 µm for the 200 and 90 ms collection times respectively, with a total time of ~4.5 minutes for a single ROI.

A measurement of the Raman spectrum of a fluorescence standard (SRM 2241, National Institute of Standards and Technology (NIST), U.S.A)[33] was taken before and after the Raman measurements of the RCF samples. This was used to correct for the decreased sensitivity of the spectrometer at higher (> 1500 cm$^{-1}$) Raman shifts as part of the data pre-processing steps explained in Section 2.5. Measurements of Raman data sets were repeated three times.

## 2.5. Raman spectral data pre-processing

Raman spectra were preprocessed using MATLAB 2019b (MathWorks, Massachusetts) to remove artifacts caused by cosmic ray interactions. Additional data pre-processing was performed to generate standardized Raman data sets.[25,26] This involved the correction of the instrument-dependent Raman intensity response, background subtraction and data normalization. Spectral intensity correction was applied using the NIST correction curve followed by background subtraction using the sensitive non-linear iterative peak-clipping (SNIP) algorithm.[31,34] The SNIP technique iteratively determines a baseline to subtract from the spectrum by finding the minimum between a given point and the average value of the outer edges of a window centered on that point. For generating the dose response and calibration curves, Raman spectra at a given dose were normalized to the intensity of the polyester Raman peak at 1614 cm$^{-1}$ from the same RCF sample as discussed in Section 3.1. Raman spectra were vector normalized for dose uncertainty estimates as indicated below in Section 3. In vector normalization, the intensity value at each Raman shift is divided by the spectrum 'norm', which is defined as the square root of the sum of the squared intensities of the spectrum.[26] This accounts for slightly different measurement conditions between spectra recorded with the same experimental parameters. This enabled the comparison of intensities of the same Raman peaks between different locations of the same film (for inherent spatial heterogeneity measurements), or between different days of measurement (for data reproducibility).

## 2.6. Dose response and calibration curves

The behavior of the Raman dose response curve appeared to be similar in nature to dose response curves obtained using optical scanners.[35] The optimal analytical function to model the calibration curve was determined by fitting the data using the least squares method[36] and is given below.

$$D_{fit} = a + bx + cx^n \qquad\qquad\qquad \text{Equation (1)}$$

where $D_{fit}$ is the estimated dose (Gy), $x$ is the Raman intensity response of a particular Raman band (width $\sim 4$ cm$^{-1}$), and $a$, $b$, $c$ and $n$ are fitting parameters. It was determined that fixing $n$ resulted in significantly lower dose estimation uncertainty. Using a custom-written code in MATLAB, calibration curves with a value of $n$ that ranged between 1 and 2 with a step of 0.001 were tested. This algorithm then selected the calibration curve with an $n$ parameter which resulted in the lowest overall dose estimation uncertainty.

### 2.7. Dose uncertainty estimates

The process of generating the calibration curve involves two main sources of uncertainties that are related to the experiment, and the curve fitting process involving the fit parameters determined during the calibration.[35] The experimental uncertainty has contributions from uncertainties in the Raman spectral intensity measurement process.

Using error propagation, the total dose estimation uncertainty $\sigma_{D_{tot}}$ for the microdosimetry system of RCF and Raman micro-spectroscopy, can be divided into experimental uncertainty $\sigma_{D_{exp}}$ and the fitting uncertainty $\sigma_{D_{fit}}$

$$\sigma_{D_{exp}}(\%) = \frac{\sqrt{(b+n\cdot c\cdot x^{n-1})^2 \cdot \sigma_x^2}}{D_{fit}} \cdot 100 \qquad \text{Equation (2)}$$

$$\sigma_{D_{fit}}(\%) = \frac{\sqrt{(x\cdot \sigma_b)^2 + (x^n \cdot \sigma_c)^2}}{D_{fit}} \cdot 100 \qquad \text{Equation (3)}$$

where $\sigma_x$ is the standard deviation of the measured Raman intensity response, $D_{fit}$ is the estimated dose and where $\sigma_b$ and $\sigma_c$ are the uncertainty determined by the curve fitting algorithm for the $b$ and $c$ parameters. The experimental and fitting uncertainties can then be combined to determine an overall dose estimation uncertainty for the calibration curve.

$$\sigma_{D_{tot}}(\%) = \frac{\sqrt{(b+n\cdot c\cdot x^{n-1})^2\cdot\sigma_x^2 + (x\cdot\sigma_b)^2 + (x^n\cdot\sigma_c)^2}}{D_{fit}} \cdot 100 \qquad \text{Equation (4)}$$

## 2.8. Assessment of experimental uncertainty

The experimental uncertainty is dependent on the uncertainty $\sigma_x$ in the measured Raman intensity [Eq (2)], which has a number of contributions. The prominent contributions to $\sigma_x$ are due to (1) the scan area inhomogeneity which is determined by the reproducibility of the optical alignment of the laser beam and the Raman signal collection path, the laser power stability, and the stability of the microscope stage during scanning; (2) the inherent microscopic inhomogeneity[19,20] in the RCF on the scale of 1 - 10 μm; (3) the macroscopic millimeter scale differences between film samples[19,20]; and (4) a smaller contribution is from the realization of dose using the ionization chamber which is ~0.6% for doses ≥ 0.5 Gy, and ~2.6% for doses < 0.5 Gy. The total experimental uncertainty $\sigma_x$ determined from the Raman intensity measurements is thus equal to the square root of the sum of the squared contributions of each of sources (1 − 4) above. The contributions of sources (2) and (3) were experimentally determined as discussed below. The uncertainty caused by source (1) due to the scan area inhomogeneity, thus could be estimated from all the other measured values.

To determine the effect of source (2) which is the uncertainty due to the inherent microscopic inhomogeneity in the film, four raster scan measurements, each with the same ROI, were performed at the corners of a 2 × 2 mm$^2$ grid in a single unirradiated RCF sample. After initial pre-processing, these data were vector normalized, and the mean Raman intensity of four different ROIs was assessed for the 696, 1445, and 2060 cm$^{-1}$ Raman peaks.

The impact on the experimental uncertainty due to source (3) the macroscopic (~mm) scale heterogeneities in the RCF samples, was investigated by determining the reproducibility of Raman measurements of the same RCF sample for a given dose by performing repeated measurements that involved repositioning of the samples. For this purpose, the Raman spectral intensities for the RCF sample irradiated at 2 Gy were analyzed from six different data sets. Raman spectra were vector normalized and the mean intensities in the ROI for the Raman peaks at 696, 1445, and 2060 cm$^{-1}$ across the six data sets were compared.

## 2.9. Varying spatial resolution to assess the Raman response

The statistical noise dominates at the microscopic length scales of the Raman intensity measurement reported here due to the inherent microscopic inhomogeneities in the RCF. In earlier work,[37] the mean length of the microcrystals in the active layer of EBT3 films was determined to be 9.4 µm with a standard deviation of 5.6 µm for the distribution, and the mean width was found to be 1.6 µm with a standard deviation of 0.3 µm for the distribution. Therefore, to reduce statistical noise, the Raman intensity was averaged over longer segments. Thus 'pixel' lengths corresponding to 10, 30, 50, 75 and 150 µm provided varying sampling spatial resolution with a smaller pixel length corresponding to higher resolution. Except for the results in Section 3.5, a nominal pixel length of 30 µm was used in the data analysis. The mean of the Raman spectra from multiple spatial locations inside a single pixel is considered to represent the Raman response from the 30 µm-long pixel.

## 3. RESULTS

### 3.1. Raman spectra of RCFs

Fig. 3(a) illustrates an example of the raw data showing the mean of 865 Raman spectra over the 150 × 100 µm$^2$ scan area for each dose including 0, 0.15, 0.25, 0.50 and 0.75 Gy. The prominent peaks in the Raman spectrum at 696, 1086, 1445 and 2058 cm$^{-1}$ are assigned to vibrations of the $\delta$ (CCC), $\nu$(C−C), $\nu$(C=C) and $\nu$(C≡C) modes in the backbone of the polyPCDA active layer. The peaks at 1186, 1216, 1240 and 1333 cm$^{-1}$ are due to the CH$_2$ wagging and twisting modes of the all-trans alkyl chains.[38] The peaks at 1614 and 1725 cm$^{-1}$ are due to the stretching vibrations of the C=C and C=O bonds in the polyester layer.[39] The difference in intensity of the Raman band at the 1614 cm$^{-1}$ peak at different doses was evaluated after applying vector normalization across the 1550-1800 cm$^{-1}$ region of the spectra containing the polyester peaks. The relative standard deviation of the vector normalized intensity values of the polyester Raman band at 1614 cm$^{-1}$ for different doses was found to be ~0.2 %. This confirms that the polyester peak is not affected by the X-ray irradiation. Raman spectra at different doses were thus normalized to the intensity of the polyester peak resulting in spectra shown in Fig. 3(b).

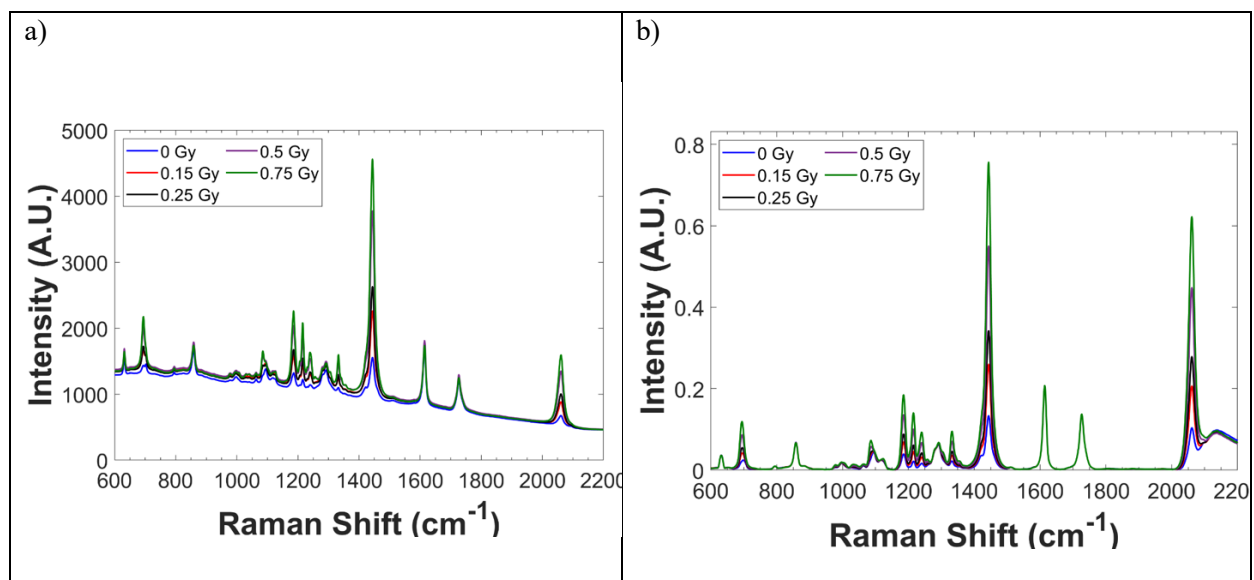

Fig. 3 (a) Raw data, showing the mean of 865 Raman spectra over the 150 x 100 μm$^2$ scan area for each dose; (b) Spectra after background correction, NIST-intensity correction and polyester peak normalization. Raman intensity is shown in arbitrary units (A.U.)

## 3.2. Dose response using the peak normalization method

After normalizing by the intensity of the polyester peak at 1614 cm$^{-1}$, the mean of the Raman spectra in the scan area of the ROI and the corresponding standard deviation was determined for each dose. The results for the change in the mean intensity of the active layer Raman peaks at 696, 1445 and 2060 cm$^{-1}$, are shown in Fig. 4 (a-c) for the low (0-2 Gy) dose range and in Fig. 4 (d-f) for the high (2–50 Gy) dose range. It is evident that the Raman peak intensities increase with dose and do not saturate even at 50 Gy and a clear difference in intensity is seen between consecutive doses in the 0-50 Gy range. The results from the student t-test also confirmed that the difference between the means of the Raman intensity for consecutive pairs of doses is highly significant (p< 0.005).

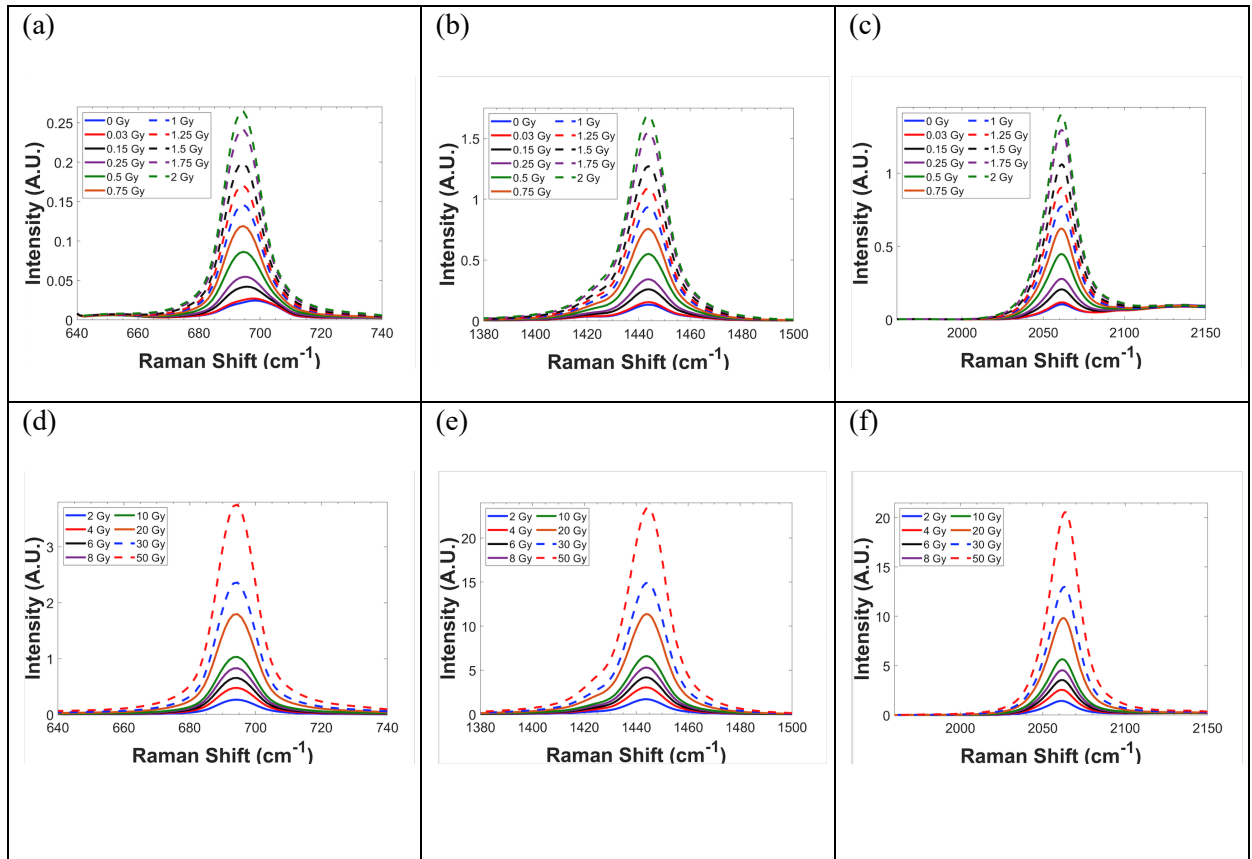

Fig. 4: Raman dose response of the RCF samples. Mean spectral intensity of three different Raman peaks in the active layer at 696, 1445 and 2060 cm$^{-1}$ for 0 - 2 Gy (a – c) and 2 – 50 Gy (d – f) dose range.

## 3.3. Dose Calibration curves

From the results for the dose response (Fig. 4), the dose delivered was plotted as a function of the measured mean intensity of the Raman peaks at 696, 1445 and 2060 cm$^{-1}$ along with error bars corresponding to the 1-sigma standard deviation. The data was fit with Eq. (1) to generate three calibration curves as shown in Fig. 5, that enable conversion of the measured Raman intensity to dose values. The best fitting parameters for the three calibration curves are in Table I.

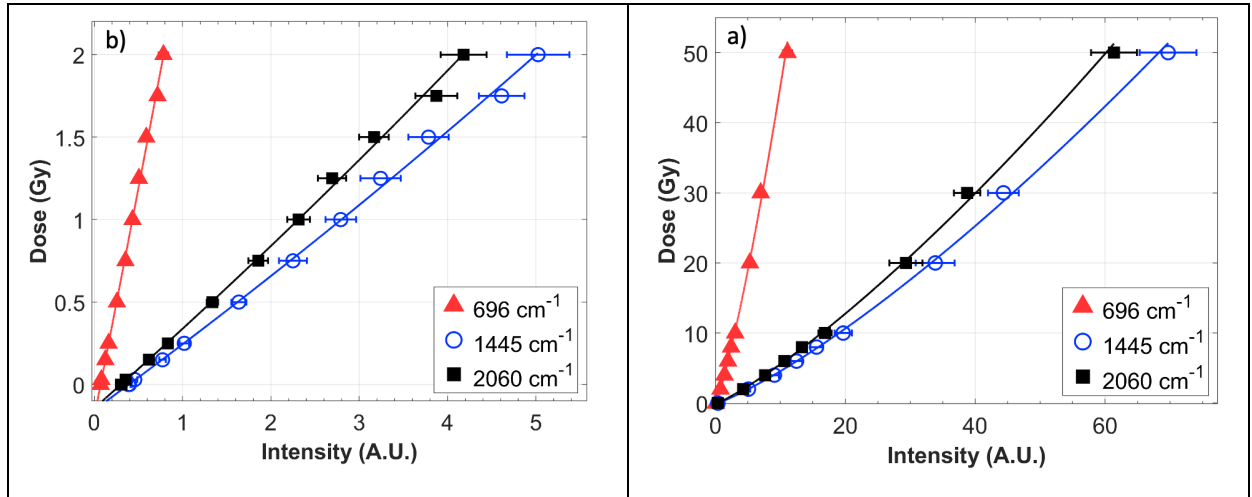

Fig. 5: Calibration curves generated using experimental data corresponding to the mean intensity values for three different Raman peaks at 696, 1445 and 2060 cm$^{-1}$ in the (a) 0-2 Gy dose range and (b) 0-50 Gy dose range. Error bars correspond to the 1-sigma standard deviation of the intensities of the specific Raman band for each sample within the ROI. For some points the error bars appear smaller than the size of the symbols.

**Table I**: Table of best-fit parameters and corresponding uncertainties for the 0-50 Gy calibration curves.

| Raman peak (cm$^{-1}$) | 'a' | $\sigma_a$ | 'b' | $\sigma_b$ | 'c' | $\sigma_c$ | 'n' |
|---|---|---|---|---|---|---|---|
| 696 | -0.170 | 0.009 | 2.12 | 0.12 | 0.762 | 0.056 | 1.50 |
| 1445 | -0.129 | 0.008 | 0.297 | 0.018 | 0.0566 | 0.0035 | 1.48 |
| 2060 | -0.154 | 0.008 | 0.398 | 0.019 | 0.0528 | 0.0036 | 1.52 |

### 3.4. Dose uncertainty estimates

The calibration curves would typically be used to determine an unknown dose to the RCF by reading off the measured Raman intensity. Thus we determined an overall uncertainty estimate in such a determination of unknown dose. Figures 6 (a), (b) and (c) display the experimental, fitting and total uncertainty calculated using Eqs (2), (3) and (4), respectively, for a representative data set, at a spatial resolution of 30 × 30 μm$^2$. Fig. 6 (c) shows that the total dose uncertainty in the 0.5 – 50 Gy dose range is fairly similar for all three Raman bands with values between 6% – 9%. The anomalous value at 20 Gy was not reproduced in the two other data sets, and was attributed to scan area inhomogeneity discussed in Section 4. The higher total dose uncertainty values of 8% - 25% for doses below 0.5 Gy is most likely due to insufficient polymerization in the EBT3 film resulting in a relatively lower sensitivity. The experimental uncertainty in the 0.5 – 50 Gy dose range is between 5% – 8% except for the 20 Gy data point as noted above. The fit uncertainty values in the 0.5 – 50 Gy dose range seem to be fairly independent of dose between 2.5% – 4.5% for all three Raman bands.

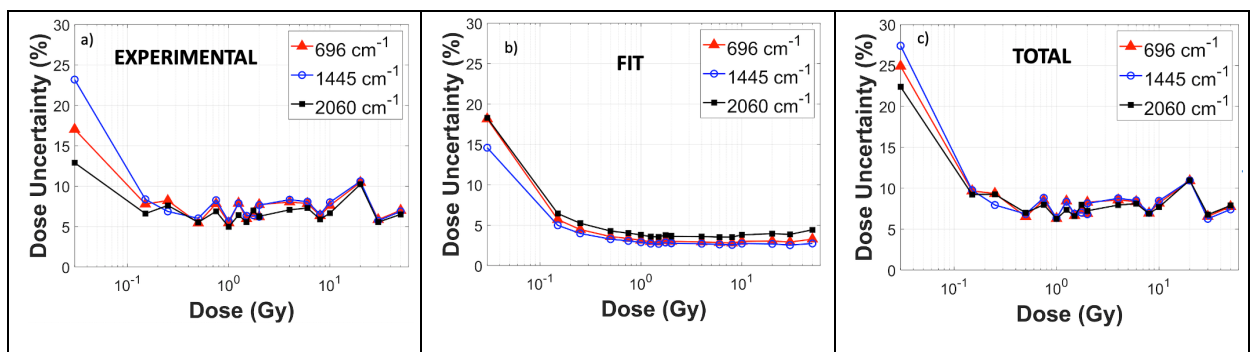

Fig. 6: (a) Experimental uncertainty (b) Fitting uncertainty and (c) Total dose measurement uncertainty for the Raman peaks at 696, 1445 and 2060 $cm^{-1}$ for the 0.03 - 50 Gy dose range.

As discussed in Section 2.8, the total experimental uncertainty in the range of 5% – 8% determined from the Raman intensity measurements has four major sources: (1) the scan area inhomogeneity during a scan; (2) the inherent microscopic film inhomogeneity; (3) the macroscopic film inhomogeneity; and (4) experimental uncertainty from the realization of dose using the ionization chamber (~0.6% for doses ≥ 0.5 Gy). The contributions of sources (2) and (3) were experimentally determined. Thus, the uncertainty caused by (1) the scan area inhomogeneity, could be estimated from all the other measured values for the total experimental uncertainty and from sources (2), (3) and (4).

The effect of the inherent microscopic inhomogeneity (source 2) on the experimental uncertainty is presented in Table S1 in supplementary materials, including the standard deviation (STD) and the relative standard deviation (RSTD) of the mean Raman intensity measured from four ROIs on a control sample.

Thus, the microscopic heterogeneity in the film contributes to an uncertainty of ~ 0.9%, 3.3% and 1.7% in the experimental dose uncertainty estimates for the 696, 1445 and 2060 cm[-1] Raman peaks. These results are consistent with dose uniformity results reported earlier using confocal Raman spectroscopy of RCF [24.]

The impact on the experimental uncertainty due to source (3) the macroscopic millimeter scale heterogeneities was measured from six sets of data and is demonstrated in Table S2 in supplementary materials. This was found to be 2.2%, 1.3% and 1.4 % for the 696, 1445 and 2060 cm[-1] Raman peaks respectively, from the relative standard deviation in the mean Raman intensity response at 2 Gy.

The different components of the uncertainties mentioned above are summarized in Table II for the read-out operation of the dose based on the 2060 cm[-1] Raman peak calibration curve as an example. The mean values of the uncertainties are listed for doses in the 0.5 – 50 Gy dose range.

**Table II**: Typical uncertainty contributions based on the 30 µm spatial resolution read-out of the 2060 cm[-1] Raman peak intensities for doses ≥ 0.5 Gy.

| Uncertainties | Nominal Value |
|---|---|
| **Experimental** (Total contribution from the measurement of Raman spectral intensities) | 6.2% |
| Measured contribution from source (2) the inherent micrometer-scale spatial heterogeneity in the RCF sample | 1.7% |
| Measured contribution from source (3) the macroscopic heterogeneity (reproducibility in the measurement of the same RCF sample) | 1.4% |
| Measured contribution from source (4) the delivery of dose to the RCF | 0.6% |
| Estimated[a] contribution caused by (1) the scan area inhomogeneity | 5.8% |
| **Fit** | 4.0% |
| **Total** | 7.4% |

[a]Estimated from the square root of (the square of experimental uncertainty from Raman intensity

measurements minus the square of each of the contributions from sources (2), (3) and (4))

### 3.5 Effect of sampling spatial resolution on the total dose uncertainty

The statistical noise due to the inherent microstructure and granularity of the PDA crystals in the EBT3 film can be reduced by considering the average response of multiple spatial points inside single 'pixels' of select sampling spatial resolution. The total dose uncertainty estimated above was for the spatial resolution of $30 \times 30~\mu m^2$ across the ROI of $150 \times 100~\mu m^2$. The total dose uncertainty was estimated for other pixel lengths of 10, 50, 75 and 150 μm and plotted along with the results for 30 μm as shown in Fig. 7. As expected, increasing the pixel length reduces the total dose uncertainty since the uncertainty due to the statistical noise reduces by inverse of the square root of the area of the pixel.[19] Similar results were found using the optical microscope read-out method of irradiated HD-V2 film.[20]

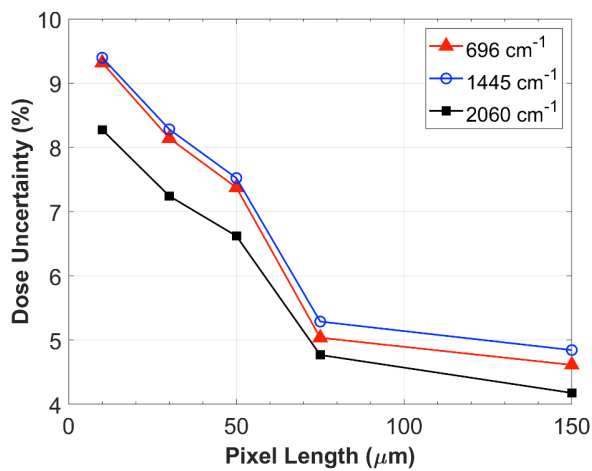

Fig. 7: Total dose uncertainty as a function of the spatial resolution (pixel length) for the film irradiated at 2 Gy.

## 4. DISCUSSION

Previous work that demonstrated the feasibility of Raman spectroscopy to read irradiated RCFs[22,24] relied on a commercial confocal Raman microscopy setup using an excitation wavelength of 632.8 nm in contrast to 785 nm used in this work. The polyester peaks around 1614 and 1725 cm$^{-1}$ were not present in the Raman spectra reported earlier.[22,24] The Raman shifts in the spectrum are relative to the laser wavelength. Hence the polyester peaks at Raman shifts of 1614 and 1725 cm$^{-1}$ should be present in the spectrum regardless of whether a red (632 nm) or near infra-red laser (785 nm) are utilized for excitation, provided that the depth of focus of the laser is large enough so that the Raman scattered light is collected concurrently from both the active layer and the polyester layer. In previous work, the depth of focus of the confocal microscope used was ~2.5 µm. This was most likely the reason why no Raman peak from the overlaying polyester layer was observed in the spectrum when the laser was focused in the active layer. A peak ratio method[24] that involved the ratio of the intensities of the Raman peaks at 2060 and 696 cm$^{-1}$ in the RCF spectrum, was applied to produce a dose response curve. The data were fit by an exponential function, and the resolution of dose response at the measured low doses of 0.3, 0.5, 1 and 2 Gy was marginal. Moreover the dose response seemed to plateau around a dose of 20 Gy or higher, which was attributed to the decrease in intensity of the 696 cm$^{-1}$ Raman peak with increased dose.

In contrast, the custom Raman micro-spectroscopy setup utilized in this work had a depth of focus of ~34 µm. This ensured that Raman scattered light from the polyester laminate was collected concurrently with that from the active layer which was only 28 µm thick. This allowed for the normalization of the collected Raman spectra to the intensity of the 1614 cm$^{-1}$ Raman peak from the polyester laminate which is unaltered by radiation. As a direct consequence of this normalization approach and contrary to results by Mirza and co-workers,[22,24] our data (Fig. 4 and Fig. 5) show high resolution in the dose response even down to 0.03 Gy in addition to an increase of the mean intensity of all three prominent Raman peaks in the RCF with increasing dose. The dynamic range for all three Raman bands was found to extend to 50 Gy with no observation of saturation. The calibration curves generated using the Raman peak response at 1445 and 2060 cm$^{-1}$ are similar in terms of the sensitivity to dose, displaying similar levels of uncertainty in the 0.5 – 50 Gy dose range as seen in Fig. 6 (c), and as such either is a viable choice for a calibration curve. In comparison, a small change in Raman intensity read out using the 696 cm$^{-1}$ calibration curve will cause a large change in the estimated dose value. Hence the 696 cm$^{-1}$ Raman band is not a good choice for the calibration curve.

The lateral resolution and depth of focus of the multimode laser spot in this work is 30 μm and 34 μm respectively with the sample moved through the laser focus using a step size of ~0.55 μm or ~0.25 μm in a raster scan pattern. This type of resolution could be useful for applications such as microbeam radiation therapy[40] or small field dosimetry.[41] For applications such as experimental microdosimetry[10] that require a smaller spatial resolution of ~1 μm, a single mode laser and a microscope objective with a larger numerical aperture (NA), along with a point-by-point mapping could be used in the setup to achieve this higher resolution. The advantageous feature of the longer depth of focus of ~34 μm for concurrent accumulation of the Raman signal from the polyester laminate, can still be retained in the high (1 μm) spatial resolution setup, by implementing an approach based on, for example, depth sensitive Raman spectroscopy.[42]

The dose values predicted using the calibrations curves in Fig. 5 are accompanied by total dose uncertainties plotted in Fig. 6, that are made up of contributions from the experimental and fit uncertainty, and evaluated using Equations (2-4). The illustrative example of the dose uncertainty budget shown in Table II for the 2060 cm$^{-1}$ Raman calibration curve, suggests that the dominant portion of the experimental uncertainty in the dosimetry system presented here arises from the heterogeneity in the measurement process during a single scan of the ROI. This can potentially be reduced by implementing a tighter control of the optical beam alignment, and by improving the stability of the microscope stage while scanning. In addition, changes that may occur in the incident laser power during a single raster scan of the ROI could be monitored and used as feedback to correct for the changes in the Raman intensity. The fitting uncertainty of ~4% reported here could be further reduced by performing Raman measurements at further intermediate dose values in the high and low dose range. The exponent 'n' in Equation (1) models the nonlinear dose response in the high dose region (> 10 Gy). In this work, 'n' was fixed between 1 and 2 while 'a', 'b' and 'c' were varied using a step-wise procedure[35,36] implemented using MATLAB and described in Section 2.6, which resulted in the lowest dose estimation uncertainty. When 'n' was a floating parameter, the limited number of dose points (at 20, 30 and 50 Gy) in the high dose region resulted in higher (by up to 88 %) total dose estimation uncertainties. Future work will investigate the use of more datapoints at higher doses as well as alternate fitting approaches to assess the uncertainties in the fit. We expect a slight increase, similar to the 1% - 2% increase reported earlier,[35,36] in the fit uncertainties. The contribution to the experimental uncertainty from the realization of dose using the ionization chamber is only 0.6% for doses $\geq$ 0.5 Gy. The uncertainty is higher (around 2.6 %) for doses < 0.5 Gy because of linac start-up effects. Further reduction of this uncertainty at low (<0.5 Gy) doses could be achieved by irradiations performed in beams from cobalt-60 or cesium-137 or kV x-ray irradiators that are well-characterized for radiation protection level doses.

The two-dimensional dosimetry system in this study was based on reading out the dose response of the EBT3 type of RCF using Raman micro-spectroscopy. A similar approach could be extended to construct and characterize high spatial resolution dosimetry systems based on Raman micro-spectroscopy read-out of other types of RCFs. For example, a system based on EBT-XD film[13] could provide Raman calibration curves with an even larger dose range and lower uncertainties at the high spatial resolution of 10 µm, while a system based on XR-QA2 film[13] could provide Raman calibration curves for the very low dose range of 1 mGy - 0.2 Gy along with potentially lower uncertainties. Such high spatial resolution experimental dosimetry systems will be useful to compare with results from computational microdosimetry approaches involving Monte Carlo simulations[9] to enable high accuracy evaluation of the energy deposition within microscopic biological cell volumes. This will benefit Raman micro-spectroscopic investigations of cellular response to ionizing radiation,[5-8] that could lead to further advances in radiation therapy.

## 5. CONCLUSIONS

This work presents a novel approach to high spatial resolution dosimetry based on Raman micro-spectroscopy read-out of RCFs. In contrast to earlier work based on confocal Raman spectroscopy measurements of RCFs, the results reported here enable the measurement of dose response in the RCF over an extended dose range from 0-50 Gy, and in particular at very low doses in the 0-2 Gy range. The radiation response of three Raman intensity bands associated with the vibrational frequencies of PDA molecules in the RCF was used to generate calibration curves that could be used to determine an unknown dose to the RCF. The two-dimensional dosimetry system was characterized for the first time, in terms of the uncertainty in the estimation of the total delivered dose at microscopic spatial resolution. The contribution to the total dose uncertainty from the calibration curve fit uncertainty and from different sources of experimental uncertainty was evaluated. The main contribution to the experimental uncertainty was from the scan area heterogeneity caused by instabilities in the optical alignment, the laser power and the stage scanning process. These could be readily controlled, and future work will focus on lowering these experimental uncertainty contributions to the evaluation of dose.

Micrometer-scale spatial resolution and sensitivity to low doses remain a challenge for radiation dosimetry. The high spatial resolution experimental dosimetry system of Raman micro-spectroscopy and RCF could potentially be used for applications in microdosimetry to produce meaningful dose-estimates in cellular targets, as well as for applications based on small field dosimetry that involve high dose gradients. This could support applications in areas of novel radiation therapy treatments with nanodevices, brachytherapy, radiosurgery, and radiation biology that require high spatial resolution dosimetry.

## ACKNOWLEDGEMENTS

The authors acknowledge the support of the Natural Sciences and Engineering Research Council of Canada (NSERC), the Canada Research Chairs program, and the Ministry of Research and Innovation of Ontario. Thanks to Rob Crohn and David Hood at Ashland Specialty Ingredients, for samples of the EBT3 radiochromic films. Thanks to Taylon Clark for help with the Raman measurements of RCF, and to Harry Allen for the pre-processing algorithms for the Raman spectra. Thanks to Dr. Brad Behr at Tornado Spectral Systems for technical support related to the spectrometer, and to Philippe Gravelle and Mike Antunes at Carleton University for the custom sample mount and the neutral density filters.

## CONFLICTS OF INTEREST

The authors have no relevant conflicts of interest to disclose.

# REFERENCES

___________________

[1] Butler HJ, Ashton L, Bird B, Cinque G, Curtis K, Dorney J, Esmonde-White K,  Fullwood NJ, Gardner B, Martin-Hirsch PL, Walsh MJ, McAinsh MR, Stone N, and Martin FL. Using Raman spectroscopy to characterize biological materials. *Nature Protocols*. 2016;11(4):664.

[2] Kong k, Kendall C, Stone N, Notingher I. Raman spectroscopy for medical diagnostics. *Advanced Drug Delivery Reviews*. 2015;89:121–134.

[3] Austin LA, Osseiran S, Evans CL. Raman technologies in cancer diagnostics. Analyst. 2016 Jan 21;141(2):476-503. doi: 10.1039/c5an01786f. PMID: 26539569.

[4] Hartmann K, Becker-Putsche M, Bocklitz T, et al. A study of Docetaxel-induced effects in MCF-7 cells by means of Raman microspectroscopy. *Anal Bioanal Chem*. 2012;403(3):745-753. doi:10.1007/s00216-012-5887-9.

[5] Matthews Q, Brolo A, Lum J, Duan X, Jirasek A. Raman spectroscopy of single human tumour cells exposed to ionizing radiation in vitro. *Phys Med Biol*. 2011;56:19–38.

[6] Harder SJ, Isabelle M, DeVorkin L, Smazynski J, Beckham W, Brolo AG, Lum JJ, Jirasek A. Raman spectroscopy identifies radiation response in human non-small cell lung cancer xenografts. *Scientific Reports*. 2016;6:21006. http://doi.org/10.1038/srep21006.

[7] Allen CH, Kumar A, Nyiri B, Qutob S, Chauhan V and Murugkar S. Raman micro-spectroscopy analysis of human lens epithelial cells exposed to a low-dose-range of ionizing radiation. *Phys Med Biol.* 2018;63:025002.

[8] Meade AD, Howe O, Unterreiner V, Sockalingum GD, Byrne HJ, Lyng FM. Vibrational spectroscopy in sensing radiobiological effects: analyses of targeted and non-targeted effects in human keratinocytes. *Faraday Discuss*. 2016 Jun 23;187:213-34.

[9] Oliver PAK, Thomson RM. Microdosimetric considerations for radiation response studies using Raman spectroscopy. Med Phys. 2018;45(10): 4734-4743.

[10] Rahmanian S, Niklas M, Abdollahi A, Jäkel O, Greilich S. Application of fluorescent nuclear track detectors for cellular dosimetry. *Phys Med Biol.* 2017;62(7):2719-2740. doi: 10.1088/1361-6560/aa56b4.

[11] Bartzsch S, Corde S , Crosbie JC et al. Technical advances in x-ray microbeam radiation therapy. *Phys. Med. Biol.* 2020; 65:02TR01-02TR28.

[12] Devic S. Radiochromic film dosimetry: Past, present, and future. *Phys Med*. 2011; 27: 122-134.

[13] Devic S, Tomic N, Lewis D. Reference radiochromic film dosimetry: Review of technical aspects. *Phys Med*. 2016;32(4):541-556. doi:10.1016/j.ejmp.2016.02.008

[14] Niroomand-Rad A, Chiu-Tsao ST, Grams MP et al. Report of AAPM Task Group 235 Radiochromic Film Dosimetry: An Update to TG-55. *Med Phys.* 2020;47(12):5985-6025.

[15] Aldelaijan S, Devic S, Papaconstadopoulos P et al. Dose–response linearization in radiochromic film dosimetry based on multichannel normalized pixel value with an integrated spectral correction for scanner response variations. *Med Phys.* 2019;46(11)5335-5350.

[16] Callens M, Crijns W, Simons V, et al. A spectroscopic study of the chromatic properties of GafChromicEBT3 films. *Med Phys.* 2016;43(3):1156-1166. doi:10.1118/1.4941312

[17] Borca VC, Pasquino M, Russo G, Grosso P, Cante D, Sciacero P, Girelli G, Rosa La Porta M, Tofani S. Dosimetric characterization and use of Gafchromic EBT3 film for IMRT dose verification. *J Appl Clin Med Phys*. 2013;14:158–171.

[18] Massillon-Jl G, Chiu-Tsao ST, Domingo-Muñoz I, Chan MF. Energy dependence of the new Gafchromic EBT3 film: Dose response curves for 50 kV, 6 and 15 MV x-ray beams. *Int J Med Phys, Clin Eng Radiat Oncol.* 2012;1:60–65.

[19] Bartzsch S, Lott J, Welsch K, Br€auer-Krisch E, Oelfke U. Micrometer-resolved film dosimetry using a microscope in microbeam radiation therapy. *Med Phys.* 2015; 42:4069–4079.

[20] Pellicioli P, Bartzsch S, Donzellia M, Krisch M, Bräuer-Krisch E. High resolution radiochromic film dosimetry: Comparison of a microdensitometer and an optical microscope. *Phys Med.* 2019;65:106–113.

[21] Casolaro P, Campajola L, Breglio G, Buontempo S, Consales M, Cusano A, Cutolo A, Di Capua F, Fienga F, Vaiano P. Real-time dosimetry with radiochromic films. *Nature Sci Rep.* 2019;9:5307.

[22] Mirza JA, Park H, Park SY, Ye SJ. Use of radiochromic film as a high-spatial resolution dosimeter by Raman spectroscopy. *Med Phys.* 2016;43(8):4520. doi:10.1118/1.4955119

[23] Talarico OS, Krylova TA, Melnik NN. Raman scattering for dosimetry using GAFCHROMIC EBT3 radiochromic dosimetry film. *Med Phys.* 2019;46:1883–1887.

[24] Mirza JA, Millares RH, Kim GI, Park SY, Lee J, Ye SJ. Characterization of radiochromic films as a micrometer-resolution dosimeter by confocal Raman spectroscopy. *Med Phys.* 2019;46(11):5238–5248.

[25] Morais, C.L.M., Paraskevaidi, M., Cui, L. et al. Standardization of complex biologically derived spectrochemical datasets. *Nat. Protoc. 2019;* 14: 1546–1577. https://doi.org/10.1038/s41596-019-0150-x

[26] Gautam R, Vanga S, Ariese F, Umapathy S. Review of multidimensional data processing approaches for Raman and infrared spectroscopy. *EPJ Tech Instrum.* 2015;2:1–38.

[27] GafchromicTM EBT3 film specifications. http://www.gafchromic.com.

[28] Seuntjens J, Olivares M, Evans M, Podgorsak E. Absorbed dose to water reference dosimetry using solid phantoms in the context of absorbed-dose protocols: Dose to water reference dosimetry using solid phantoms. *Med Phys.* 2005;32(9):2945–2953. doi: 10.1118/1.2012807.

[29] McEwen M, DeWerd L, Ibbott G, et al. Addendum to the AAPM's TG-51 protocol for clinical reference dosimetry of high-energy photon beams. *Med Phys.* 2014;41(4):041501.

doi:10.1118/1.4866223

[30] McEwen MR. Measurement of ionization chamber absorbed dose kQ factors in megavoltage photon beams: Measurement of photon kQ factors. *Med Phys*. 2010 Apr;37(5):2179–2193. doi: 10.1118/1.3375895.

[31] Hansson B, Allen CH, Qutob S, Behr B, Nyiri B, Chauhan V, Murugkar S. Development of a flow cell based Raman spectroscopy technique to overcome photodegradation in human blood. *Biomed Opt Express*. 2019;10(5):2275-2288.

[32] Tornado Spectral Systems. Our Advantage. https://tornado-spectral.com/spectroscopy/our-advantage. Accessed February 28, 2021.

[33] Choquette SJ, Etz ES, Hurst WS, Blackburn DH, Leigh SD. Relative Intensity Correction of Raman Spectrometers: NIST SRMs 2241 Through 2243 for 785 nm, 532 nm, and 488 nm/514.5 nm Excitation. *Appl Spectrosc*. 2007;61(2):117-129.

[34] Morhac M, Matousek V. Peak Clipping Algorithms for Background Estimation in Spectroscopic Data. *Appl Spectrosc*. 2008;62(1):91-106.

[35] Devic S, Seuntjens J, Hegyi G, Podgorsak EB, Soares CG, Kirov AS, et al. Dosimetric properties of improved GafChromic films for seven different digitizers. *Med Phys*. 2004;31:2392e401.

[36] Devic S, Seuntjens J, Sham E, Podgorsak EB, Kirov AS, Schmidtlein RC, et al. Precise radiochromic film dosimetry using a flat-bed document scanner. *Med Phys*. 2005;32:2245e53.

[37] Callens MB, Crijns W, Depuydt T, Haustermans K, Maes F, D'Agostino E, Wevers M, Pfeiffer H, Van Den Abeele K. Modeling the dose dependence of the vis-absorption spectrum of EBT3 GafChromic™ films. *Med Phys*. 2017;44:2532-2543. doi:10.1002/mp.12246.

[38] Callens M, Crijns W, Simons V, De Wolf I, Depuydt T, Maes F, Haustermans K, D'hooge J, D'Agostino E, Wevers M, Pfeiffer H, Van Den Abeele K, A spectroscopic study of the chromatic properties of Gafchromic™ EBT3 films. *Med Phys*. 2016;43:1156–1166.

[39] Cho LL. Identification of textile fiber by Raman microspectroscopy. *Forensic Science Journal*. 2007;6(1):55-62.

[40] Ghita M, Fernandez-Palomo C, Fukunaga H, Fredericia PM, Schettino G, Bräuer-Krisch E, Butterworth KT, McMahon SJ, Prise KM. Microbeam evolution: from single cell irradiation to pre-clinical studies. *Int J Radiat Biol*. 2018 Aug;94(8):708-718. doi: 10.1080/09553002.2018.1425807.

[41] Palmans H, Andreo P, Huq MS, Seuntjens J, Christaki KE, Meghzifene A. Dosimetry of small static fields used in external photon beam radiotherapy: Summary of TRS-483, the IAEA-AAPM international Code of Practice for reference and relative dose determination. *Med Phys*. 2018 Nov;45(11):e1123-e1145. doi: 10.1002/mp.13208.

[42] Liu W, Ong YH, Yu XJ, Ju J, Perlaki CM, Liu LB, Liu Q. Snapshot depth sensitive Raman spectroscopy in layered tissues. *Opt Express* 2016;24:28312-28325.